\documentclass{article}
\usepackage{ijcai26}

\usepackage{times}
\usepackage{soul}
\usepackage{url}
\usepackage[hidelinks]{hyperref}
\usepackage[utf8]{inputenc}
\usepackage[small]{caption}
\usepackage{graphicx}
\usepackage{amsmath}
\usepackage{amsthm}
\usepackage{booktabs}
\usepackage{algorithm}
\usepackage{algorithmic}
\usepackage[switch]{lineno}

\usepackage{amssymb}
\usepackage[utf8]{inputenc}
\usepackage[T1]{fontenc}    
\usepackage{hyperref}       
\usepackage{url}    
\usepackage[table,dvipsnames]{xcolor}
\usepackage{booktabs}      
\usepackage{amsfonts}      
\usepackage{nicefrac}       
\usepackage{microtype}       
\usepackage{graphicx}
\usepackage{booktabs}
\usepackage{tabularx}
\usepackage{array}
\usepackage{amsmath}
\usepackage{arydshln} 
\usepackage{makecell}
\usepackage{multirow}
\usepackage{colortbl}
\definecolor{gray20}{gray}{0.9}
\definecolor{gray40}{gray}{0.8}
\usepackage{ragged2e}
\newcolumntype{L}[1]{>{\raggedright\arraybackslash}p{#1}}

\usepackage{tabularx}

\usepackage{xurl}
\usepackage{stfloats}
\usepackage{circledsteps}
\usepackage{amsthm}
\usepackage[skins]{tcolorbox}
\tcbuselibrary{breakable}
\usepackage{pifont}

\usepackage{amsthm}

\theoremstyle{plain}

\newtheorem{agent}{Agent}
\newtheorem{goal}{Goal}

\newtcolorbox[auto counter, number within=section]{prompt}[3][]{%
  enhanced,
  breakable,
  colback=#2!5!white,
  colframe=#2!75!black,
  title=\textbf{Box \thetcbcounter: #3},
  fontupper=\normalsize\fontfamily{cmss}\selectfont,
  #1
}

\newtcolorbox{questionbox}{
enhanced,
boxrule=0pt,frame hidden,
borderline west={4pt}{0pt}{Emerald!75!black},
colback=Emerald!9!white,
sharp corners
}

\newtcolorbox{nerbox}{
  enhanced,
  boxrule=0pt,
  frame hidden,
  borderline west={3pt}{0pt}{black!40},
  colback=black!3,
  sharp corners,
  fontupper=\linespread{0.75},
  left=4pt,
  right=4pt,
  top=3pt,
  bottom=3pt
}

\newtcolorbox{taskbox}{
  enhanced,
  boxrule=0pt,
  frame hidden,
  borderline west={3pt}{0pt}{SkyBlue!55!black},
  colback=SkyBlue!9!white,
  sharp corners,
  fontupper=\linespread{0.85}\selectfont,
  left=3pt,
  right=3pt,
  top=2pt,
  bottom=2pt
}

\title{An LLM-based Chain-of-Response Counter-Scam System}

\author{
Heedou Kim$^{1,2}$\and
Mogan Gim$^3$\and
Donghee Choi$^4$\and
Hoonick Lee$^1$\and
Soonil Bae$^5$\and
Mi-Young Kim$^{6,*}$\And
Jaewoo Kang$^{1,*}$\\
\affiliations
$^1$ Korea University\\
$^2$ Korean National Police Agency\\
$^3$ Hankuk University of Foreign Studies\\
$^4$ Pusan National University\\
$^5$ Korean National Police University\\
$^6$ University of Alberta\\
$^*$ Co-corresponding authors\\
\emails
\{heedou123, hoonick, kangj\}@korea.ac.kr,
gimmogan@hufs.ac.kr,
dchoi@pusan.ac.kr,
soonil.bae@gmail.com,
miyoung2@ualberta.ca
}

\begin{document}

\maketitle

\begin{abstract}
    The rapid evolution of online scams, driven by transnational networks and mass-produced social engineering scenarios, has exposed the speed limitations of conventional detection, necessitating tighter inter-agency coordination. While LLMs show promise in scam identification, their role in accelerating integrated response frameworks remains underexplored. We propose Counter-Scam, a unified LLM-based multi-agent framework that orchestrates end-to-end response from initial detection to crime investigation. The framework first proposes safe data guidelines, emphasizing non-public scam data and secure dataset construction via scam-specific NER. Developed with insights from 37 stakeholders to reduce delays and improve analytical efficiency, the system integrates CSRA (multi-agent mitigation), CSRT (nine role-aligned NLP tasks), and CSRD (a corpus of 185,300 scam cases and 38,587 knowledge entries). Experiments show that fine-tuned sLLMs surpass commercial models by over 10\% in all CSRT tasks and a 0.24 F1 improvement in scam-specific NER. This proves the framework's capability for enabling rapid, collaborative mitigation of online scam.
\end{abstract}

\def\scam{\textbf{\textsc{Counter-Scam}} }
\section{Introduction}

Online scams have become a serious global threat driven by transnational criminal networks, causing substantial financial losses of innocent people and increasing the burden on law enforcement~\cite{interpol}. Scammers employ sophisticated social engineering scenarios across calls, messaging apps, and social media by impersonating family members, romantic partners, financial institutions, or authorities, inflicting severe psychological harm on victims~\cite{fbi,yosiandra2024unveiling}. In this context, LLM-based agents specialized in social engineering content show strong potential as next-generation scam response technologies by enabling real-time detection and delivering warnings to victims~\cite{shen2025warned,kumarage2025personalized}.

\begin{figure}[t]
    \centering
    \includegraphics[width=1\columnwidth]{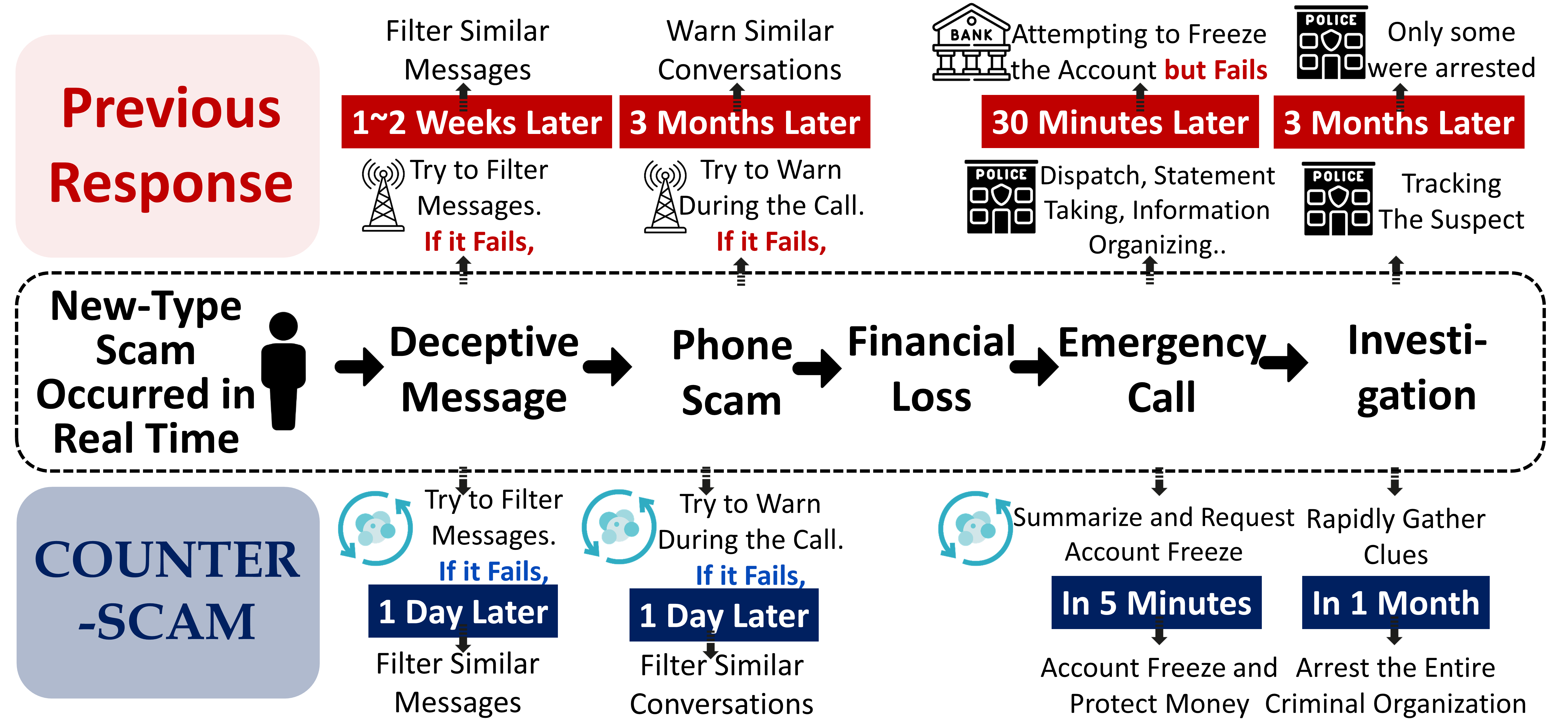} 
    \caption{Previous scam response systems often fail to keep pace with the rapid fund extraction of criminals due to fragmented inter-agency coordination and manual analysis. \scam~enables efficient end-to-end scam response.}
    \label{fig:fig1}
\end{figure}

However, the practical effectiveness of these technologies is often limited by critical delays in institutional response. As shown in Figure~\ref{fig:fig1}, when a citizen receives a new type of smishing involving a fake payment, delays in blacklisting allow scammers to complete their deception. Even with emergency reports, account freezing often lags behind the speed of illicit withdrawals. These systemic delays, stemming from inter-agency coordination and manual analysis, span detection, reporting, and intervention, enabling transnational criminal networks to outpace defenses. An integrated framework is urgently needed to enable rapid, coordinated multi-agency responses and close these gaps~\cite{tundis2019role}.

Despite the agentic potential of LLMs, most existing studies remain detection-centric and fail to address coordination delays across institutions and jurisdictions. Even with accurate detection, fragmented downstream responses can lead to failure, forcing citizens to navigate disconnected systems for actions such as account freezing and investigation and often missing the critical window for financial recovery~\cite{westmore2022crossSectorScams,luong2024understanding}. This limitation is particularly severe as international criminal networks exploit weakly regulated jurisdictions and real-time payment systems to accelerate cross-border fund exfiltration~\cite{Mitchell2025CambodiaScam}.

We propose \textbf{\scam}, an LLM-based multi-agent framework for end-to-end online scam response. It proactively blocks or detects scam-related content in real time across channels such as phone calls and text messages and, when prevention fails, rapidly reports relevant information to law enforcement agencies and security teams, integrating prevention, emergency response, and investigation. By enabling multiple stakeholders to jointly leverage the agentic capabilities of LLMs, the framework supports rapid inter-agency sharing and analysis to prevent scams and minimize harm.

\textbf{\scam} is a unified framework designed to integrate the previously fragmented stages of scam response organically, including scam detection, blocking, emergency intervention, and post-incident handling. To achieve this, it is composed of \textbf{C}hain-of-\textbf{S}cam-\textbf{R}esponse \textbf{A}gents (\textbf{CSRA}), which explicitly model the full lifecycle of scam response through specialized agents; \textbf{C}hain-of-\textbf{S}cam-\textbf{R}esponse \textbf{T}asks (\textbf{CSRT}), a suite of NLP tasks that guide LLMs toward role-aligned decision-making for each agent; and \textbf{C}hain-of-\textbf{S}cam-\textbf{R}esponse \textbf{D}ataset (\textbf{CSRD}), a carefully constructed dataset grounded in real-world scam response knowledge to ensure safe, reliable, and effective task execution. To the best of our knowledge, this work is the first to move beyond \textit{standalone} scam detection and present an \textit{integrated} agent–task–data framework that covers the \textit{entire} online scam response.

\textbf{CSRA} was developed through interviews with experts in cybersecurity and criminal investigation, resulting in an agent-based framework for comprehensive phone and text based scam response. It organizes mitigation into three stages including pre scam, immediate post victimization, and repeated victimization, guided by a situational analysis agent that determines appropriate actions. Three specialized agents then support execution: a scam prevention agent for proactive blocking and real time detection, an emergency response agent for immediate interventions such as reporting and account freezing, and an investigation agent for large scale analysis of organized scam activities. By covering the full scam cycle and linking information across stages, \textbf{CSRA} reduces response gaps and enables coordinated mitigation. Through this, it aims to complement law enforcement capacities and strengthen collaboration against international scams.

For practical deployment, we adopted strict safety principles. To prevent misuse, we release only non-scam data, apply scam-specific NER–based de-identification, and actively involve domain experts. Under this framework, the construction of \textbf{CSRT} and \textbf{CSRD} proceeds by (1) defining additional eight tasks that capture the core capabilities of each agent based on prior work, and (2) collecting original scam cases and scam response knowledge, including legal and operational materials, from the Korean National Police to build expert-annotated training and evaluation datasets. As a result, we construct a large-scale corpus comprising 185,300 scam cases and 38,587 response knowledge entries, providing a strong foundation for comprehensive scam response agents.

Under real-world constraints where commercial models are restricted, we fine-tuned sLLMs on the \textbf{CSRD}. Our results show that scam-specific NER consistently outperformed commercial models by 0.24 F1, and the fine-tuned sLLMs exceeded GPT-4o and Gemini’s average performance by 10\%, showing improvements of up to 0.55 in deceptive message analysis compared to zero-shot baselines. Nonetheless, complex reasoning tasks such as legal analysis and long-context emergency report summarization remain challenging, highlighting the need for further training. These results demonstrate the potential and effectiveness of \scam as an integrated scam response system and point toward directions for further enhancement. The main contributions are:

\begin{enumerate}
    \item We propose \textbf{\scam}, the first integrated agent–task–data framework covering the entire cycle of scams for an end-to-end response.
    \item Involving 37 stakeholders, we developed \textbf{CSRA} (agents), \textbf{CSRT} (tasks), and \textbf{CSRD} (misuse-resistant data) for practical LLM-based scam response.
    \item We demonstrate that fine-tuned sLLMs significantly outperform commercial large models on scam-response-specific 9 tasks.
    \item By releasing our framework and large datasets,\footnote{\href{https://github.com/Heedou/multiagent_scam_response}{Github repository}}, we provide a foundation for scalable deployment and cross-border collaboration in scam mitigation.
\end{enumerate}

\section{Related Work}

\subsection{Evolution of Online Scams and Challenges}
Online scams exploit human psychological vulnerabilities through social engineering, using SMS, calls, and social media to impersonate family, acquaintances, recruiters, or authorities and defraud victims~\cite{kumarage2025personalized,ai2024defending}. Transnational criminal networks based in Southeast Asia, such as Cambodia and Vietnam, orchestrate these schemes, exploiting legal gaps and money-laundering networks~\cite{amnesty2025cambodia}. Effective mitigation requires coordinated cross-border and cross-institution responses, yet fragmented law enforcement and weak collaboration allow scams to persist~\cite{shen2025warned,loggen2022unraveling}.

\subsection{LLM-based Scam Response}
Language models have emerged as crucial tools in scam response, capable of accurately identifying scams and providing interpretable explanations~\cite{kim-etal-2026-scriptmind,koide2024chatspamdetector,lee2024korsmishing}. Recent advances in multimodal processing and real-time detection have extended their utility beyond mere plausibility checks, enabling more realistic and practical applications in real-world settings~\cite{cao2025phishagent,shen2025warned}. Building on this trend, LLMs are increasingly envisioned as agent-based automated tools that can assist human responders at critical intervention points, supporting faster and more effective scam mitigation~\cite{xue2025multiphishguard,afane2024next}.

\begin{figure*}[h!]
    \centering
    \includegraphics[width=0.75\textwidth]{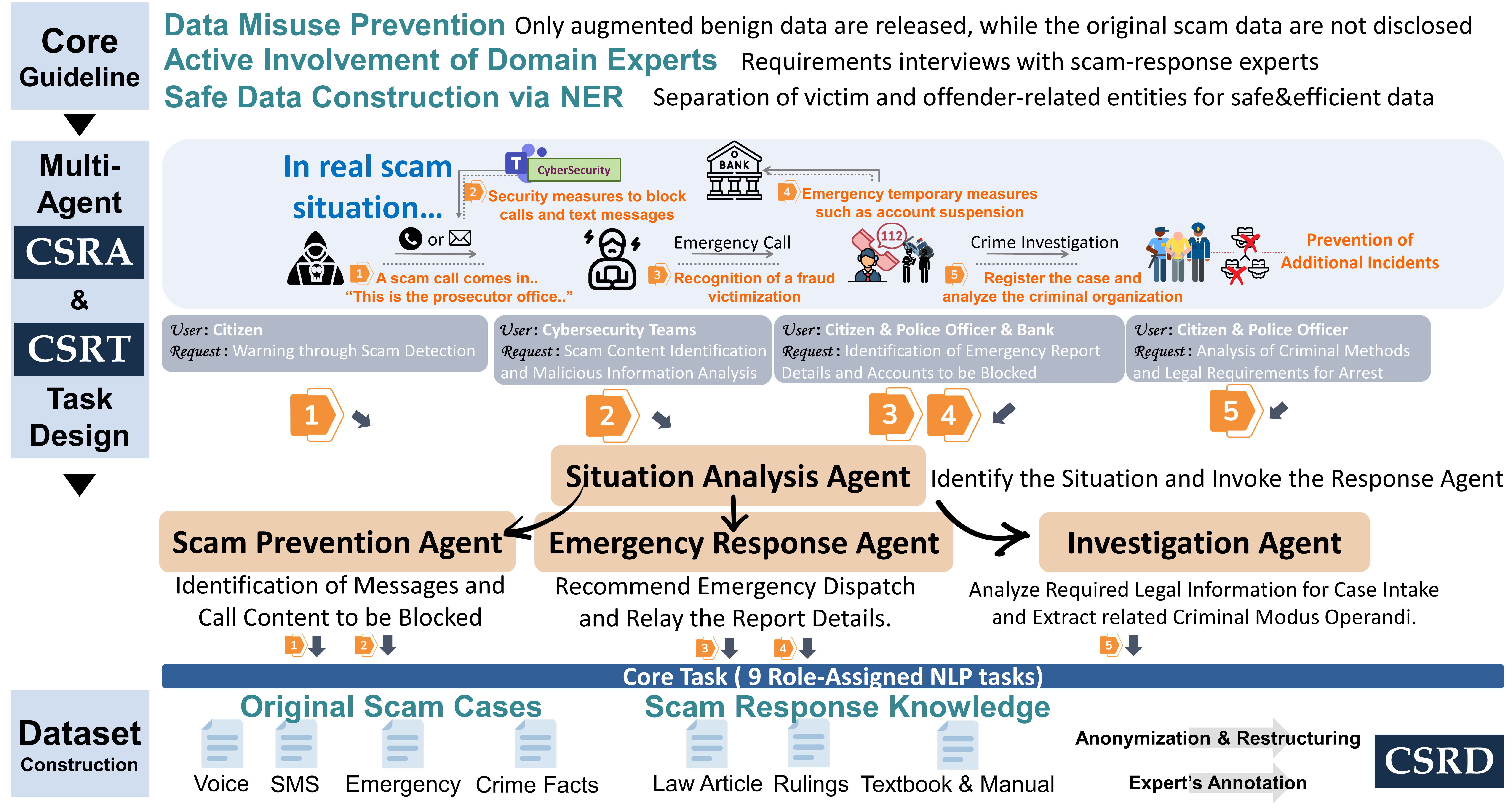} 
    \caption{\scam is an agent-based system that, unlike prior LLM studies focusing solely on scam content detection, supports AI-driven coordinated scam response across citizens, security teams, law enforcement, and banks within the real-world scam response workflow.}
    \label{fig:overview}
\end{figure*}

\section{Method}

\begin{table*}[h!]
\centering
\footnotesize
\setlength{\tabcolsep}{3pt}
\renewcommand{\arraystretch}{0.7}
\begin{tabular}{p{2.7cm} p{3.5cm} p{3.8cm} p{1.7cm} p{1.9cm} p{2.1cm} p{3.5cm}}
\hline
Agent & Task & Purpose & Cases & Knowledge & Instance Type \\
\hline

All Agents &
Case Analysis NER &
Crime entity extraction
(malicious links,
suspect numbers) &
$D_{Phone}$: 14k \newline
$D_{SMS}$: 25k &
-- &
$\{X_{Tok}, Y_{NER}\}$ \\

\hline

\multirow{2}{*}{Situation Analysis} &
Offense Detection &
Crime type
identification &
$D_{Crime}$: 144k &
-- &
$\{X_{Cri}, Y_{Cri}\}$ \\
\cline{2-7}

&
Operational QA &
Response planning &
QA &
$K_{PM}$: 5.6k &
$\{X_{Q}, Y_{A}\}$ \\

\hline

\multirow{2}{*}{Scam Prevention} &
Scam Scenario
Detection &
Scam conversation
detection &
$D_{Phone}$: 14k &
$K_{Benign}$: 14k &
$\{X_{Scam}, Y_{Scam}\}$ \\
\cline{2-7}
&
Message
Classification &
Fraud modus
classification &
$D_{SMS}$: 25k &
-- &
$\{X_{Sms}, Y_{M.O.}\}$ \\

\hline

Emergency Response &
Call
Summarization &
Rapid victim
report summary &
$D_{ER}$: 2.3k &
-- &
$\{X_{Em}, Y_{Em}\}$ \\

\hline

\multirow{3}{*}{Investigation} &
Hypothesis
Eval &
Determination of criminality with its rationales &
-- &
$K_{CI}$: 687 \newline
$K_{CL}$: 3.3k \newline
$K_{CR}$: 15k &
$\{X_{Hyp}, Y_{Hyp}\}$ \\
\cline{2-7}
&
Statute Mapping &
Applicable law
prediction &
$D_{Crime}$: 144k &
-- &
$\{X_{Cri}, Y_{Law}\}$ \\
\cline{2-7}
&
Element Analysis &
Legal element
identification &
$D_{Crime}$: 144k &
-- &
$\{X_{Cri}, Y_{Com}\}$ \\

\hline
\end{tabular}
\caption{The purpose, type and dataset formats of all tasks (\textbf{CSRT}) assigned to the \scam Agent (\textbf{CSRA}).}
\label{tab:msrt_ultra_compact}
\end{table*}

Our key insight is that, just as large-scale scam organizations operate through internal coordination and role specialization, effective scam response requires seamless collaboration among all stakeholders including government, victims, and industry. As illustrated in Figure~\ref{fig:overview}, this integration can be realized through a multi-agent framework. Based on interviews with field experts, we derived design principles for such agents. The resulting \textbf{CSRA} defines clear roles and responsibilities for each agent, \textbf{CSRT} structures LLM-based NLP tasks to operationalize these roles, and \textbf{CSRD} provides a safe and reliable dataset for task execution and evaluation. We further validate the framework’s practical feasibility through evaluations.

\subsection{Establishing Core Guidelines}

A key requirement for an effective scam response ecosystem involving multiple stakeholders is the seamless integration of incident data with real-time information. In practice, however, the inherent risks and potential for misuse of scam data have led to restricted access, which in turn has hindered the development of integrated response systems. To address this gap, we propose core guidelines for designing datasets that ensure safety while enabling efficient collaboration among stakeholders, and empirically demonstrate their implementation.

\paragraph{\ding{172} Data Misuse Prevention} We adopt a principle of not releasing real scam conversation data that could be exploited by fraudsters for scenario generation or victim information theft~\cite{permana2023personal}. Instead, we provide carefully curated benign data, enabling the development of detection models while minimizing false positives. Additionally, officially produced texts such as emergency report dialogues, legal summaries of criminal incidents, and communications between law enforcement and citizens are made available to support the generation of appropriate response actions.

\paragraph{\ding{173} Active Involvement of Domain Experts} We actively engaged domain experts to develop the practical components necessary for implementing \scam, spanning agent design, task definition, and dataset creation. A total of 37 experts from four fields including general police officers, criminal investigators, cyber fraud investigators and policy analysts participated in deriving agent designs, evaluating the operational suitability of proposed tasks through preliminary interviews, and performing data labeling and quality assessments. The reliability of their annotations and evaluations was ensured through consistent adherence to classification guidelines, supplemented by inter-annotator agreement measurements and correlation analyses to validate the results.

\paragraph{\ding{174} Safe Data Construction via NER} For efficient and safe scam response, we designed the \textsc{Case Analysis NER}. This pipeline clearly distinguishes victim information, perpetrator details, and impersonated identities, minimizing exposure of sensitive data while preserving analytically meaningful signals for crime analysis. In particular, it is designed to extract impersonated identities (e.g., fake names) and criminal tools (e.g., malicious links and account numbers), enabling their use for crime prevention and tracking~\cite{kim2022crimeNER}. We constructed a dataset from 28,806 texts including phone scam conversations, smishing messages, and crime fact data, labeled with 10 crime-domain and 13 general-domain entity types using BIO tagging.

\begin{nerbox}
\label{task:case_analysis_ner}
\textbf{\textcolor{Black!55!black}{\small \textsc{Case Analysis NER}}} \\
Let $D = \{X_{\text{Tok}}, Y_{\text{NER}}\}$. For a scam-related text, each token sequence $X_{\text{Tok}} \in X$, where $X \subseteq \{D_{\text{Phone}}, D_{\text{Sms}}, D_{\text{Crime}}\}$, maps to a BIO-labeled entity sequence $Y_{\text{NER}}$ identifying crime-specific (e.g., fake identities, tools) and general entities (e.g., dates).
\end{nerbox}

\subsection{LLM-Based Multi-Agent(CSRA)}

\begin{goal}[End-to-End Scam Response]
The objective of \scam is to identify the phase of an online scam and deliver phase-appropriate response to the relevant stakeholders in a timely manner. Since online scams unfold as lifecycle-oriented crimes rather than single events, effective mitigation requires structured, phase-aware responses aligned with specific analytical goals and responsible actors. We define an agent \(A = \langle R, G, I, O, S \rangle\) by its role \(R\), objective \(G\), input space \(I\), output space \(O\), and stakeholders \(S\).
\end{goal}

\begin{agent}[Situation Analysis]
\label{agent:situationanalysis}
The Situation Analysis Agent \(A_{\mathrm{SA}}\) is designed to assess scam situations across heterogeneous users, where \(R_{\mathrm{SA}}\) denotes situation assessment, \(G_{\mathrm{SA}}\) is to identify the user’s current situation and analytical needs, and \(I_{\mathrm{SA}}\) consists of scam messages, emergency reports, victim summaries, and contextual metadata such as user type and message timing. Given an input \(i \in I_{\mathrm{SA}}\), the agent produces an output \(o \in O_{\mathrm{SA}}\) through joint reasoning over input content and metadata as \(o = A_{\mathrm{SA}}(i)\), and invokes the downstream agent. Stakeholders are categorized into four groups:

\begin{itemize}
  \item \textbf{Citizens}: targets of scam attacks
  \item \textbf{Cybersecurity teams}: analyze blacklists and develop detection models to proactively block scams
  \item \textbf{Police officers}: respond to scams through emergency actions and investigations
  \item \textbf{Banks}: handle financial losses resulting from scam-related fund transfers
\end{itemize}

\end{agent}

\begin{agent}[Scam Prevention]
\label{agent:crimeprevention}

The stakeholders \(S_{\mathrm{SP}}\) include \textit{citizens} and \textit{cybersecurity teams}. The role \(R_{\mathrm{SP}}\) aims to prevent scam-related harm via early detection. Given an input message \(i \in I_{\mathrm{SP}}\), the agent generates an output \(o \in O_{\mathrm{SP}}\) by identifying scam, issuing warnings to citizens, or, for security teams, classifying social engineering techniques and extracting malicious indicators.

\end{agent}

\begin{agent}[Emergency Response]
\label{agent:emergencyresponse}

The stakeholders \(S_{\mathrm{ER}}\) are \textit{citizens}, \textit{police}, and \textit{banks}. The role \(R_{\mathrm{ER}}\) supports urgent scam response and emergency intervention. Given an input \(i \in I_{\mathrm{ER}}\) consisting of urgent victim reports and contextual metadata, the agent outputs \(o \in O_{\mathrm{ER}}\) by generating an emergency-focused summary for immediate protective action, forwarding it to relevant law enforcement agencies, and, for police officers, extracting action-critical entities and coordinating with financial institutions and telecom providers.

\end{agent}

\begin{agent}[Investigation]
\label{agent:investigator}
The stakeholders \(S_{\mathrm{I}}\) are \textit{citizens}, \textit{police}. The role \(R_{\mathrm{I}}\) supports post-incident investigation and legal reasoning. Given an input \(i \in I_{\mathrm{I}}\) consisting of detailed victim accounts and contextual information, the agent outputs \(o \in O_{\mathrm{I}}\) by assessing criminal liability, generating a statute-grounded legal analysis, and extracting impersonated identities and instrumentalities to trace recurring crimes linked to the same criminal organization.

\end{agent}

\subsection{Agent-Specific Core Tasks(CSRT)}

We define NLP tasks that underpin each agent’s decision-making capabilities. Guided by a review of prior work on scam detection and crime-related NLP, we identified tasks critical for these capabilities and categorized the types of textual inputs that stakeholders may provide during the scam response process. The task inputs include criminal facts \(X_{\mathrm{Cri}}\), scam response-related queries \(X_{\mathrm{Que}}\), conversations suspected of being scam-related \(X_{\mathrm{Scam}}\), scam messages \(X_{\mathrm{Sms}}\), general texts containing crime-related information \(X_{\mathrm{Tok}}\), victims' urgent statements \(X_{\mathrm{Em}}\), and criminal hypotheses \(X_{\mathrm{Hyp}}\).

\paragraph{\ding{172} Situation Analysis Agent} Just as police officers rely on foundational procedural knowledge to make situational judgments~\cite{holgersson2008police}, the Situation Analysis Agent must accurately assess the state of scam victimization and select the appropriate response pathway among prevention, emergency, and investigation.

\begin{taskbox}
\label{task:operational_qa}
\textbf{\textcolor{Emerald!55!black}{\small \textsc{Offense Detection}}} $D = \{X_{\text{Cri}}, Y_{\text{Cri}}\}$. For a description of suspected criminal activity, each input $X_{\text{Cri}} \in X$, where $X \subseteq \{D_{\text{Crime}} \}$, maps to a label $Y_{\text{Cri}}$ corresponding to one of 111 offense categories. \\ \\
\textbf{\textcolor{Emerald!55!black}{\small \textsc{Operational QA}}} $D = \{X_{\text{Que}}, Y_{\text{Ans}}\}$. For procedural questions related to scam response, each input $X_{\text{Que}} \in X$, where $X \subseteq \{K_{\text{PM}}\}$, maps to an answer $Y_{\text{Ans}}$ derived from police procedural manuals.

\end{taskbox}
\noindent The agent can first determine whether a conduct constitutes a criminal offense and identify its basic offense category through \textsc{Offense Detection}, and then use \textsc{Operational QA} to decide the appropriate next steps.

\paragraph{\ding{173} Scam Prevention Agent} To prevent imminent scams, the Scam Prevention Agent must detect content in real time, warn citizens, and filter malicious links and phone numbers extracted from previously identified scam cases of the same scenario to block their redistribution.

\begin{taskbox}
\label{task:fradulent_scenario_detection}
\textbf{\textcolor{Emerald!55!black}{\small \textsc{Scam Scenario Detection}}} $D = \{X_{\text{Scam}}, Y_{\text{Scam}}\}$. For a given conversation, $X_{\text{Scam}} \in X$, where $X \subseteq \{D_{\text{Phone}}, K_{\text{Benign}}\}$, maps to a label $Y_{\text{Scam}}$ indicating whether the input is fraudulent or benign.
\\ \\
\textbf{\textcolor{Emerald!55!black}{\small \textsc{Scam Message Classification}}} $D = \{X_{\text{Sms}}, Y_{\text{Sms}}\}$. For SMS messages, $X_{\text{Sms}} \in D_{\text{Sms}}$ represents bait messages by scammers to deceive victims, and maps to a label $Y_{\text{Sms}}$ indicating one of seven scam tactics.
\end{taskbox}

\noindent The agent can identify conversations in real time and issue warnings to consenting citizens, while classifying scam tactics in reported or large-scale analyzed texts and, through \textsc{Case Analysis NER}, extracting malicious indicators to be blocked and providing them to cybersecurity teams.

\paragraph{\ding{174} Emergency Response Agent} When a victim has transferred funds due to a scam, immediate action such as prompt reporting and emergency account freezing by financial institutions is essential to mitigate losses.

\begin{taskbox}
\label{task:emergency_report_summarization}
\textbf{\textcolor{Emerald!55!black}{\small \textsc{Emergency Call Summarization}}} \\
$D = \{X_{\mathrm{Em}}, Y_{\mathrm{Em}}\}$. For an emergency report, input $X_{\mathrm{Em}} \in D_{\mathrm{E.R.}}$ consists of transcripts of emergency calls, and maps to a summary $Y_{\mathrm{Em}}$ that appropriately reflects the situation.
\end{taskbox}

\noindent The agent supports urgent responses by police and financial institutions through rapid summarization of reports following the occurrence of scam victimization. The summarized report facilitates dispatch, and, together with account extracted via \textsc{Case Analysis NER}, enables prompt account freezing.

\paragraph{\ding{175} Investigation Agent} Just as a crime investigator collects evidence and pursues offenders~\cite{roberts2012law,stainton2025criminal}, the Investigator Agent extracts criminal methods from texts, including impersonation details and instruments like accounts and phone numbers, to trace repeat offenders and support prosecution under relevant legal statutes.

\noindent The agent supports investigations of large-scale scam cases to prevent further harm. It ensures lawful investigation by assessing criminality (\textsc{Criminal Hypothesis Evaluation}), mapping applicable laws (\textsc{Statute Mapping}), and identifying statutory elements (\textsc{Element Analysis}), thereby supporting clear investigative reports. Information extracted by \textsc{Case Analysis NER} further enables case clustering and follow-up investigation. 

\begin{taskbox}
\label{task:criminal_hypothesis}
\textbf{\textcolor{Emerald!55!black}{\small \textsc{Criminal Hypothesis Evaluation}}} \\$D = \{{X_{\text{Hyp}}, Y_{\text{Hyp}}}\}$. Each input $X_{\text{Hyp}} \in X$, derived from ${K_{\text{CI}}, K_{\text{CR}}}$ and reformulated as a hypothesis requiring legal assessment, maps to an output $Y_{\text{Hyp}}$ consisting of a binary judgment (true or false) and a corresponding rationale.
\end{taskbox}

\begin{taskbox}
\label{task:criminal_hypothesis}
\textbf{\textcolor{Emerald!55!black}{\small \textsc{Statute Mapping}}} $D = \{{X_{\text{Cri}}, Y_{\text{Law}}}\}$. Given a crime description $X_{\text{Cri}}$, , each input $X_{\text{Cri}} \in X$, where $X \subseteq \{D_{\text{Crime}} \}$, the task maps it to one or more applicable criminal law articles, where $Y_{\text{Law}}$ is selected from 160 statutorily defined provisions.\\ \\
\textbf{\textcolor{Emerald!55!black}{\small \textsc{Element Analysis}}} $D = \{{X_{\text{Cri}}, Y_{\text{Com}}}\}$. Given a crime description $X_{\text{Cri}}$, each input $X_{\text{Cri}} \in X$, where $X \subseteq {D_{\text{Crime}}}$, is mapped to one or more legal elements constituting the offense, where $Y_{\text{Com}}$ is selected from 129 statutorily defined components extracted from law articles.
\end{taskbox}

\subsection{Dataset Construction}
We built benchmark datasets for all tasks through data processing and expert annotation. Data were collected from two sources, Original Scam Cases for detection and Response Knowledge for situational and legal reasoning.

\paragraph{Original Scam Cases} As summarized in Table~\ref{tab:msrt_ultra_compact}, the Original Scam Case data comprise four subsets: \(D_{\mathrm{Phone}}, D_{\mathrm{Sms}}, D_{\mathrm{E.R.}},\) and \(D_{\mathrm{Crime}}\). \(D_{\mathrm{Phone}}\) includes 14,030 real voice phishing conversations collected from the Korean National Police Agency, substantially larger and more realistic than prior datasets~\cite{boussougou2024korean,ma2025teleantifraud}. \(D_{\mathrm{Sms}}\) consists of 24,947 victim-reported scam SMS messages, representing a significantly larger real-world smishing corpus than existing benchmarks~\cite{timko2024smishing,tanbhir2025bangla}. \(D_{\mathrm{E.R.}}\) is built from 2,361 anonymized call transcripts between victims and police officers absent from public dispatch datasets~\cite{vera911,sf911}. Finally, \(D_{\mathrm{Crime}}\) contains descriptions of criminal victimization derived from legal benchmarks and supplemented with synthetic cases~\cite{hwang2022multi,KLAID}.

\paragraph{Scam Response Knowledge} Effective scam response requires not only domain-specific knowledge of scams but also alignment with general law-enforcement protocols and legal frameworks. Accordingly, our knowledge base is designed to cover comprehensive crime-response knowledge and is organized into five components: \(K_{\mathrm{PM}}, K_{\mathrm{Benign}}, K_{\mathrm{CI}}, K_{\mathrm{CL}},\) and \(K_{\mathrm{CR}}\). \(K_{\mathrm{PM}}\) captures 5,660 police procedural knowledge, constructed from internal manuals spanning 380 crime types and 57 response domains. \(K_{\mathrm{Benign}}\) is introduced to reduce false positives in scam detection by augmenting dialogue data that reflect legitimate police-initiated calls to citizens, based on real investigator examples. \(K_{\mathrm{CI}}, K_{\mathrm{CL}},\) and \(K_{\mathrm{CR}}\) provide criminal investigation, criminal law, and judicial precedent knowledge, respectively, leveraging the publicly available LAPIS knowledge base~\cite{kim2024lapis}.

\paragraph{Expert-Annotated Task Dataset} For \textsc{Operational QA}, questions and answers were generated from \(K_{\mathrm{PM}}\) and reviewed by police experts, resulting in 2,624 high-quality QA pairs. Unlike prior legal QA benchmarks~\cite{zhong2020jecqa,goebel2024overview}, our benchmark captures real-world, manual-grounded police procedures. For \textsc{Offense Detection} and \textsc{Statute Mapping}, offense labels assigned to all 143,893 \(D_{\mathrm{Crime}}\) instances were consolidated using the standard charge classification table~\cite{police2023manual}, mapping them to 111 unified offense categories and corresponding legal labels. For \textsc{Scam Scenario Detection}, we constructed balanced benign counterparts to \(D_{\mathrm{Phone}}\) by collecting non-fraud dialogues from financial consultations~\cite{AIhub} and the Korean Dialogue Corpus~\cite{NIKL}, and incorporating simulated official calls from \(K_{\mathrm{Benign}}\). For \textsc{Scam Message Classification}, three crime analysts annotated all messages with one of seven fraud tactic categories. For \textsc{Emergency Call Summarization}, experts summarized dual-perspective summaries (caller and officer) from anonymized \(D_{\mathrm{E.R.}}\) transcripts. For \textsc{Element Analysis}, we analyzed 160 legal provisions and extracted 129 common crime elements, covering both conduct and result.

\subsection{Evaluation under Real-World Constraints}
We evaluated whether \scam can operate under real-world constraints such as closed-network environments, limited computation, and the unavailability of commercial LLM APIs. We compared commercial LLMs, including chatgpt-4o-latest and gemini-2.0-flash, as performance upper bounds against deployable open-source sLLMs across our tasks and datasets, including a 1B-scale model to test extreme resource limits. Multilingual sLLMs such as Llama-3.1-8B-Instruct, SOLAR-10.7B-Instruct, and Llama-3.2-1B-Instruct were evaluated for deployment on police intranet servers or on-device police phones, alongside a Korean-tuned model, EEVE-Korean-Instruct-10.8B, for Korean crime-response settings. Commercial models were evaluated in zero-shot settings, while sLLMs were tested under zero-shot inference, domain-specific fine-tuning, and RAG, with fine-tuning performed for five epochs using QLoRA and the Paged AdamW optimizer on two A100 80GB GPUs.

\section{Experiment Results}

\paragraph{Metrics} For \textsc{Operational QA}, to reflect deployment constraints where commercial models cannot access police manuals, we applied RAG-based QA prompts only to open-source models. For classification tasks, we used rule-based post-processing to correct format errors and align outputs with predefined labels~\cite{xia2024fofo}, and evaluated performance using Accuracy and F1 score. For open-ended tasks such as summarization and QA, we adopted an LLM-as-a-Judge framework~\cite{chiang2023can}, using expert-annotated gold references and instructing the judge model to evaluate outputs strictly against those references~\cite{zheng2023judging}. Prior studies comparing expert and automatic evaluations on police manuals have demonstrated the validity of this approach~\cite{lee2026evaluating}, and in our study, correlations with human evaluations exceeded 0.8, confirming strong agreement with experts. 

\subsection{Crime-domain NER is effective, yet human validation is still essential for safety.}

\begin{table}[t]
\centering
\small
\setlength{\tabcolsep}{1pt}
\renewcommand{\arraystretch}{0.6}
\begin{tabular}{lcccccc}
\toprule
\textbf{Entity} & \textbf{Gemini} & \textbf{GPT4} & \textbf{EEVE} & \textbf{Llama8B} & \textbf{Llama1B} & \textbf{SOLAR} \\
\midrule

\multicolumn{7}{l}{\textbf{Crime Domain}} \\

fake position   & .39 & .63 & \textbf{.70} & \cellcolor{gray!25}\textbf{.88} & .52 & \textbf{.72} \\
accused name    & .22 & .28 & \textbf{.82} & \cellcolor{gray!25}\textbf{.92} & .70 & \textbf{.81} \\
case number     & .22 & .76 & \textbf{.83} & \cellcolor{gray!25}\textbf{.92} & .76 & .79 \\
account         & .14 & .35 & .09 & \cellcolor{gray!25}\textbf{.94} & \textbf{.59} & .03 \\
product         & .08 & \textbf{.44} & .21 & \cellcolor{gray!25}\textbf{.72} & .32 & .12 \\
bad link        & .11 & .25 & \textbf{.43} & \cellcolor{gray!25}\textbf{.54} & .34 & \textbf{.48} \\
fakename        & .38 & .38 & \textbf{.54} & \cellcolor{gray!25}\textbf{.72} & .49 & .38 \\
crime tool      & .38 & .61 & \textbf{.74} & \cellcolor{gray!25}\textbf{.90} & .62 & .55 \\
law             & .22 & .24 & \textbf{.69} & \cellcolor{gray!25}\textbf{.83} & .49 & \textbf{.71} \\
fake org        & .23 & .18 & \textbf{.48} & \cellcolor{gray!25}\textbf{.64} & \textbf{.52} & .51 \\

\midrule

\multicolumn{7}{l}{\textbf{General Domain}} \\

relation        & .15 & .27 & \textbf{.75} & \cellcolor{gray!25}\textbf{.86} & .60 & .66 \\
birth           & .49 & \textbf{.65} & \textbf{.59} & \cellcolor{gray!25}\textbf{.75} & .45 & .29 \\
identity        & .00 & .24 & \textbf{.33} & \textbf{.33} & .30 & \cellcolor{gray!25}\textbf{.36} \\
age             & .32 & .40 & \textbf{.76} & \cellcolor{gray!25}\textbf{.97} & .54 & .59 \\
location        & .32 & .28 & \textbf{.51} & \cellcolor{gray!25}\textbf{.67} & .48 & .37 \\
datetime        & .31 & .64 & \textbf{.82} & \cellcolor{gray!25}\textbf{.94} & \textbf{.81} & .79 \\
occupation      & .22 & .28 & \textbf{.47} & \cellcolor{gray!25}\textbf{.58} & .33 & .41 \\
price           & .35 & .35 & .52 & \cellcolor{gray!25}\textbf{.94} & \textbf{.76} & .38 \\
organization    & .38 & \textbf{.43} & \textbf{.48} & \cellcolor{gray!25}\textbf{.64} & .42 & .42 \\
plate           & \textbf{.63} & \cellcolor{gray!25}\textbf{.78} & .50 & \textbf{.84} & .29 & .50 \\
transport       & \textbf{.66} & .52 & .52 & \cellcolor{gray!25}\textbf{.95} & .58 & .13 \\
name            & .40 & .61 & \textbf{.72} & \cellcolor{gray!25}\textbf{.92} & \textbf{.76} & .80 \\
position        & .04 & .10 & .29 & \cellcolor{gray!25}\textbf{.80} & .19 & \textbf{.29} \\

\bottomrule
\end{tabular}

\caption{Experiment Results of \textsc{Case Analysis NER}.}
\label{tab:case_analysis_ner_compact}

\end{table}

\begin{table*}[h!]
\centering
\footnotesize
\renewcommand{\arraystretch}{0.65}
\begin{tabular}{p{3.5cm}lcccccc}
\toprule
\multirow{2}{*}{\textbf{Task}} & \multirow{2}{*}{\textbf{Metric}} & 
\multirow{2}{*}{\textbf{GPT-4o}} & 
\multirow{2}{*}{\textbf{Gemini 2.0}} & 
\textbf{EEVE} & \textbf{SOLAR} & \textbf{Llama8B} & \textbf{Llama1B} \\
 &  &  &  & \textbf{SFT} & \textbf{SFT} & \textbf{SFT} & \textbf{SFT} \\
\midrule
\multicolumn{8}{l}{\textit{\textbf{Situation Analysis}}} \\ 
Operational QA & LLM Judge & 0.69 & 0.66 & \textbf{0.87} & 0.85 & \textbf{0.88} & 0.64 \\  \cline{2-8}
\multirow{2}{*}{Offense Detection} & ACC & 0.86 & 0.86 & \textbf{0.87} & \textbf{0.98} & 0.50 & 0.21 \\
 & F1 & 0.9 & 0.93 & \textbf{0.95} & \textbf{0.99} & 0.77 & 0.61 \\ 
\midrule
\multicolumn{8}{l}{\textit{\textbf{Scam Prevention}}} \\ 
\multirow{2}{*}{\begin{tabular}[c]{@{}l@{}}Scam Scenario Detection\end{tabular}} & ACC & 0.97 & 0.87 & \textbf{0.99} & \textbf{0.99} & 0.86 & 0.63 \\
 & F1 & 0.97 & 0.88 & \textbf{0.99} & \textbf{0.99} & 0.85 & 0.58 \\ \cline{2-8}
\multirow{2}{*}{\begin{tabular}[c]{@{}l@{}}Scam Message Classification\end{tabular}} & ACC & 0.88 & 0.93 & \textbf{0.97} & \textbf{0.99} & \textbf{0.97} & 0.88 \\
 & F1 & 0.7 & 0.76 & 0.91 & \textbf{0.98} & \textbf{0.95} & 0.73 \\ 
\midrule
\multicolumn{8}{l}{\textit{\textbf{Emergency Response}}} \\ 
\begin{tabular}[c]{@{}l@{}}\scriptsize Emergency Call Summarization\end{tabular} & LLM Judge & \textbf{0.89} & \textbf{0.75} & 0.62 & 0.56 & 0.51 & 0.2 \\ 
\midrule
\multicolumn{8}{l}{\textit{\textbf{Investigation}}} \\ 
\multirow{2}{*}{\begin{tabular}[c]{@{}l@{}}Criminal Hypothesis Eval\end{tabular}} & ACC & \textbf{0.73} & 0.62 & \textbf{0.74} & 0.62 & 0.62 & 0.62 \\
 & F1 & \textbf{0.79} & 0.77 & \textbf{0.79} & 0.77 & 0.77 & 0.77 \\ \cline{2-8}
\multirow{2}{*}{Statute Mapping} & ACC & 0.43 & 0.4 & \textbf{0.86} & \textbf{0.88} & 0.19 & 0.07 \\
 & F1 & 0.65 & 0.69 & \textbf{0.92} & \textbf{0.95} & 0.35 & 0.12 \\ \cline{2-8}
\multirow{2}{*}{Element Analysis} & ACC & 0.67 & \textbf{0.81} & 0.66 & \textbf{0.71} & 0.64 & 0.12 \\
 & F1 & 0.84 & \textbf{0.93} & 0.81 & \textbf{0.88} & 0.82 & 0.24 \\
\bottomrule
\end{tabular}
\caption{Comparative Performance Evaluation of Fine-tuned sLLMs and Commercial LLMs}
\label{main-result}
\end{table*}

As shown in Table~\ref{tab:case_analysis_ner_compact}, fine-tuned sLLMs for \textsc{Case Analysis NER} consistently outperformed commercial models, achieving substantially higher F1 scores when both B- and I-tokens were correctly identified. While commercial models averaged an F1 of 0.35 across all entity tokens, fine-tuned sLLMs reached 0.59, with the best-performing models showing a large gap (GPT-4o 0.42 vs. Llama8B 0.79). In the crime domain, newly defined entities such as fake position, accused name, account, and crime tools were recognized far more accurately by fine-tuned sLLMs, indicating that even small models can effectively support safe, partially automated crime data processing. Commercial models also struggled with precise span detection for general entities, reflecting limited adaptability to crime-specific texts. However, fine-tuned sLLMs still showed weaknesses in extracting highly variable malicious elements such as bad links (maximum F1 0.54), suggesting the continued need for human review.

\subsection{A scam-specific labeled dataset is essential for reliable detection.}

\begin{table}[t]
\renewcommand{\arraystretch}{0.5}
\centering
\small
\setlength{\tabcolsep}{3pt}

\begin{tabular}{p{2.9cm}ccccc}
\toprule
\textbf{True Label} & \textbf{Inst.} & \textbf{GPT4} & \textbf{Gemini} & \textbf{EEVE} & \textbf{SOLAR} \\
\midrule

\multicolumn{6}{l}{\textbf{Scam Scenario Detection}} \\

Non Scam
& 1426
& 38
& \cellcolor{gray!25}\textbf{351}
& 4
& 4 \\

\midrule

\multicolumn{6}{l}{\textbf{Scam Message Classification}} \\

Impersonation of acquaintances
& 1845
& \cellcolor{gray!25}\textbf{3}
& 2
& 1
& 1 \\

Job recruitment fraud
& 197
& \cellcolor{gray!25}\textbf{7}
& 4
& 4
& 6 \\

Overseas delivery scam
& 100
& 15
& \cellcolor{gray!25}\textbf{16}
& 7
& 7 \\

Impersonation of institutions
& 415
& 3
& 3
& 5
& \cellcolor{gray!25}\textbf{6} \\

Voice spam
& 50
& \cellcolor{gray!25}\textbf{50}
& \cellcolor{gray!25}\textbf{50}
& 21
& 0 \\

Overseas payment fraud
& 501
& \cellcolor{gray!25}\textbf{327}
& 162
& 5
& 10 \\

Illegal loan offers
& 638
& 36
& 40
& \cellcolor{gray!25}\textbf{66}
& 10 \\

\bottomrule
\end{tabular}

\caption{False Negatives by Task and Class.}
\label{tab:fn_by_task_class}

\end{table}

Accurate detection and fine-grained classification of scams are essential, as they determine downstream actions such as warnings and blocking. Although commercial LLMs achieve high performance in \textsc{Scam Scenario Detection} (0.97 and 0.87), Table~\ref{tab:fn_by_task_class} indicates that non-negligible false negatives persist. Notably, Gemini misclassified 351 non-scam messages as scams, which can undermine system reliability. In contrast, the fine-tuned sLLM misclassified only 4 out of 1,426 instances, demonstrating substantially greater stability. The \textsc{Scam Message Classification} task further reveals consistent misclassification patterns across commercial models. Both GPT-4o and Gemini frequently confused specific scam types, with errors concentrated in \textit{Overseas payment fraud} and \textit{Voice spam}. Such errors limit the system’s ability to provide clear grounds for scam intervention by security teams.

\subsection{While fine-tuning is highly effective in the scam response task, more fine-grained training is still required.}

\paragraph{Effect of Fine-tuning sLLMs} We compared the zero-shot performance of sLLMs with their fine-tuned counterparts on 150 small samples for each task. On average across all sLLMs, fine-tuning improved performance by 0.26, and the ability to follow task-specific instructions also increased by 0.28, demonstrating a clear benefit of fine-tuning. In particular, the \textsc{Scam Message Classification} showed an average improvement of 0.55 compared to zero-shot, confirming that fine-tuning is an effective approach.

\paragraph{Fine-tuned sLLMs vs Commercial Models} Table~\ref{main-result} shows performance across all tasks. Fine-tuned sLLMs outperformed commercial LLMs in 6 out of 9 tasks, highlighting their strong adaptability to the scam response. In particular, fine-tuned EEVE 10.8B and SOLAR 10.7B achieved an average score of 0.85 and 0.87, exceeding commercial LLMs by 10\%.

\paragraph{Risk Without Scam Response Knowledge} Without useful manual knowledge enhanced by \(K_{\mathrm{PM}}\), scam response systems may hallucinate legal facts or provide operationally misleading answers, even when factually correct. We present concrete examples in the supplementary material. For instance, when asked how to delete harmful content, a commercial model merely provides a single deletion link, whereas the manual-based sLLM explains the actual procedure used in practice, including that reporting speed does not differ between individual and institutional submissions. This shows that commercial models can sometimes hinder efficient response.

\paragraph{Limitations from a Practical Usability Perspective} Despite improvements, some tasks still fall short of practical performance levels. \textsc{Operational QA} achieved only scores in the 0.8 range even with manual-based RAG, indicating the need for additional evaluation metrics such as logical completeness. \textsc{Emergency Call Summarization} also underperformed compared to commercial models despite fine-tuning, suggesting that smaller models lack the ability to effectively process long-context text. Finally, although tasks like \textsc{Element Analysis} and \textsc{Criminal Hypothesis Evaluation} showed significant gains, they still do not reach mature levels of accuracy, highlighting the need for more refined training methods for complex reasoning tasks.

\section{Conclusion}
We propose \scam, the first multi-agent framework to integrate fragmented online scam response processes. By establishing a comprehensive agent-task-data model, we extend the scope of LLMs beyond simple detection to include blocking, emergency intervention, and post-incident investigation. Leveraging a large-scale dataset of over 185,300 scam cases and 38,587 legal knowledge entries, we demonstrated that fine-tuned sLLMs outperform commercial models in scam-specific NER and all tasks, proving their practical utility. While challenges remain in complex reasoning tasks such as legal analysis, this work confirms the potential of AI agents as practical partners in scam mitigation. Through open-source deployment, we aim to provide a foundation for cross-border collaboration against global scams.

\section*{Acknowledgment}
This work was supported by the Institute of Information \& Communications Technology Planning
\& Evaluation (IITP) grant funded by the Korea government (MSIT) (No. 2022-0-00653, Development
of a Voice Phishing Information Collection, Processing, and Big Data–Based Investigation Support
System).

\bibliographystyle{named}
\bibliography{ijcai26}

\end{document}